\documentclass[aps,prb,reprint,superscriptaddress,showpacs]{revtex4-1}
\pdfoutput=1
\usepackage{color}
\usepackage{graphicx}
\usepackage{amsmath}
\usepackage{amsfonts}
\usepackage{amssymb}
\usepackage{float}
\usepackage{hyperref}

\newcommand{\ket}[1]{|#1\rangle}
\newcommand{\braket}[2]{\langle #1|#2\rangle}
\allowdisplaybreaks
\usepackage{multirow}
\usepackage{braket}
\usepackage{algorithm}
\usepackage[noend]{algpseudocode}

\usepackage[normalem]{ulem}

\begin{document}
\author{Pieter~W.\ Claeys}
\email{PieterW.Claeys@UGent.be}
\affiliation{Institute for Theoretical Physics, University of Amsterdam, Science Park 904, 1098 XH Amsterdam, The Netherlands}
\affiliation{Center for Molecular Modeling, Ghent University, Technologiepark 903, 9052 Ghent, Belgium}
\affiliation{Department of Physics and Astronomy, Ghent University, Proeftuinstraat 86, 9000 Ghent, Belgium}

\author{Jean-S\'ebastien~Caux}
\affiliation{Institute for Theoretical Physics, University of Amsterdam, Science Park 904, 1098 XH Amsterdam, The Netherlands}

\author{Dimitri~Van~Neck}
\affiliation{Center for Molecular Modeling, Ghent University, Technologiepark 903, 9052 Ghent, Belgium}
\affiliation{Department of Physics and Astronomy, Ghent University, Proeftuinstraat 86, 9000 Ghent, Belgium}

\author{Stijn~De~Baerdemacker}
\affiliation{Center for Molecular Modeling, Ghent University, Technologiepark 903, 9052 Ghent, Belgium}
\affiliation{Department of Physics and Astronomy, Ghent University, Proeftuinstraat 86, 9000 Ghent, Belgium}
\affiliation{Department of Inorganic and Physical Chemistry, Ghent University, Krijgslaan 281 (S3), 9000 Ghent, Belgium}

\title{A variational method for integrability-breaking Richardson-Gaudin models}

\begin{abstract}

We present a variational method for approximating the ground state of spin models close to (Richardson-Gaudin) integrability. This is done by variationally optimizing eigenstates of integrable Richardson-Gaudin models, where the toolbox of integrability allows for an efficient evaluation and minimization of the energy functional. The method is shown to return exact results for integrable models and improve substantially on perturbation theory for models close to integrability. For large integrability-breaking interactions, it is shown how (avoided) level crossings necessitate the use of excited states of integrable Hamiltonians in order to accurately describe the ground states of general non-integrable models.

\end{abstract}


\maketitle


\section{Introduction}

Integrable models take a special place within the broader class of quantum many-body systems\cite{korepin_quantum_1993,caux_remarks_2011,dukelsky_colloquium:_2004,gaudin_bethe_2014}. They can be solved exactly in polynomial time, and can as such be used to investigate the physics of strongly-correlated systems beyond the reach of conventional methods \cite{dickhoff_many-body_2005}. Whereas exact diagonalization of the Hamiltonian by definition also returns exact results for arbitrary systems, it is necessarily limited to small system sizes due to the exponential scaling of the Hilbert space. However, the advantage of exact solvability comes at a price - for a model to be integrable, and thus exactly solvable, all parameters and interactions of the system need extraordinary fine-tuning. Even slight perturbations to the Hamiltonian break integrability, and it is still an open question how much of the features of integrability are retained for systems `close to integrability'. Theoretically, we immediately lose the full underlying framework, and it is in general no longer possible to solve such systems exactly.

It is then a natural question to ask how well the wave functions of systems close to integrability can be approximated using exact eigenstates of integrable systems. Given a set of such trial states, we propose to perform a variational optimization in order to obtain the optimal approximation to the ground state of a given Hamiltonian within this set of eigenstates of integrable models. The main requirement for any variational method to be feasible is being able to efficiently and accurately calculate and minimize the energy functional 
\begin{equation}\label{int:en}
E\left[\psi\right] = \frac{\braket{\psi|\hat{H}|\psi}}{\braket{\psi|\psi}},
\end{equation}
for any given Hamiltonian $\hat{H}$ and any given trial state $\ket{\psi}$ \cite{sakurai_modern_2010}. The theoretical and numerical toolbox of integrability provides us with exactly this \cite{slavnov_calculation_1989,amico_exact_2002, zhou_superconducting_2002,links_algebraic_2003,faribault_exact_2008,amico_bethe_2012,faribault_determinant_2012,claeys_eigenvalue-based_2015}. The Bethe ansatz structure of these eigenstates indeed allows for a calculation of expectation values and overlaps at a computationally favorable (polynomial) scaling.

For Hamiltonians close to integrability, we can then expect the trial states to be able to capture the physics of the problem and this variational method should return an accurate approximation to the exact ground state. Thanks to the use of eigenstates of unperturbed integrable models as trial states, the variational energy is also guaranteed to be an improvement upon the energy obtained from first-order perturbation theory, serving as further motivation for the choice of trial states. More specifically, we describe a variational method using the eigenstates of Richardson-Gaudin (RG) integrable Hamiltonians \cite{richardson_restricted_1963,richardson_exact_1964,gaudin_bethe_2014} as trial states. We apply this method to spin systems consisting of an integrable model plus an integrability-breaking perturbation term. We focus on the specific class of RG integrable models because it provides us with a large amount of variational parameters\cite{dukelsky_colloquium:_2004,ortiz_exactly-solvable_2005,amico_bethe_2012} and because RG models are known to qualitatively describe a wide variety of physical systems \cite{ibanez_exactly_2009,rombouts_quantum_2010,dunning_becbcs_2011,el_araby_order_2014,ortiz_many-body_2014,van_raemdonck_exact_2014,ortiz_what_2016,claeys_read-green_2016,jurco_quantum_1989,dukelsky_exactly_2004,ortiz_exactly-solvable_2005,coish_hyperfine_2004,faribault_spin_2013,faribault_integrability-based_2013,van_den_berg_competing_2014,dukelsky_exactly_2011,de_baerdemacker_probing_2013}. The freedom in the choice of variational parameters is subsequently expected to provide accurate approximations to the ground states of a variety of non-integrable Hamiltonians. We exploit that the energy of a given Hamiltonian can be efficiently evaluated as a sum of determinants, and apply a gradient descent method to minimize the energy functional \cite{press_numerical_2007}.

If no integrability-breaking terms are present, the proposed method leads to the exact ground state by construction. Otherwise, the improvement compared to perturbation theory is investigated, and it is shown that this method is also able to return accurate approximations in the region where perturbation theory is not expected to hold, provided the perturbative interactions do not influence the qualitative physics of the model. In this case, the bulk of the correlations in the ground state of the non-integrable system is captured by the ground state of the integrable system, and the variational optimization returns an accurate approximation. If this is not the case, we show that a more accurate description can be obtained by variationally optimizing an excited state of an integrable model. This is illustrated by comparing overlaps and correlation functions, and can be understood as (avoided) level crossings in the spectrum of the non-integrable Hamiltonian\cite{dalessio_quantum_2016}.

In a broader context, this research fits within the general development of wave function-based methods (as compared to density-based methods) for the description of strongly-correlated models. In this aspect, the present study is also motivated by recent developments in the theory of Antisymmetric Product of Geminals (APG) in molecular physics and quantum chemistry \cite{surjan_introduction_1999,johnson_size-consistent_2013,limacher_new_2013,tecmer_assessing_2014}. Composed as a generalized Valence-bond wavefunction, APG wavefunctions are tailor-made for the description of resonating electron-pair configurations, and tune in directly with the Lewis picture of molecular bonding.  Notwithstanding these sound physical foundations, APG theory is severely limited for applications in large molecular systems, due to the highly multi-reference character of its wavefunction.  Recently, it has been realized that the Richardson-Gaudin eigenstates fit within the class of geminal wavefunctions.  This has given rise to various computationally tractable versions of APG, including a variational formulation based on the RG wavefunctions \cite{johnson_model_2015,tecmer_assessing_2014}. However, pioneering calculations for simple molecular systems \cite{tecmer_assessing_2014} showed that the variational method was surpassed in accuracy and efficiency by Coupled-Cluster-based APG methods \cite{limacher_new_2013,boguslawski_efficient_2014}. These preliminary results then naturally shifted the research focus to the Coupled-Cluster variant of APG theory in recent years \cite{boguslawski_efficient_2014,boguslawski_nonvariational_2014,limacher_simple_2014,henderson_seniority-based_2014, stein_seniority_2014,degroote_polynomial_2016}. However, it is presently becoming clear that further developments in APG theory will benefit from a well-defined Hilbert space, which is conveniently obtained through the connection with a variational Richardson-Gaudin APG state and the associated integrable Hamiltonian.

In principle, this method can be applied to arbitrary Hamiltonians, but in this work we focus on Hamiltonians consisting of an integrable part and a perturbative term, where the approximations can be made clearer and the advantages and limitations of the integrable wave functions can be better understood. So, one of the purposes of this work is to reassess the variational APG procedure based on the RG eigenstate and shed light on the variational procedure and eigenstate optimization. For this, the direct link with integrable systems is crucial, hence the preference for a study of a couple of minimal integrable models and integrability-breaking systems.

The paper is organized as follows. Section \ref{sec:bi} contains a discussion on breaking integrability, placing this work within the general context of integrability-based techniques for non-integrable systems. Section \ref{sec:met} presents an overview of relevant results for Richardson-Gaudin (RG) models and describes the proposed method. This is then applied to two classes of non-integrable systems in Section \ref{sec:res}, where the accuracy of the method is assessed by comparing with results from exact diagonalization for select systems. Section \ref{sec:concl} is then reserved for concluding remarks.

\section{Moving away from integrability}
\label{sec:bi}
A rich variety of methods has been developed for the approximation of the ground state of general non-integrable systems. Here, the distinction can be made between wave function-based methods such as mean-field theory \cite{chaikin_principles_2000}, the related coupled cluster and configuration interaction theories \cite{helgaker_molecular_2014}, tensor networks \cite{orus_practical_2014} and variational quantum Monte Carlo methods \cite{sorella_wave_2005} compared to density-based density functional theory \cite{parr_density-functional_1994}. Within the wave function-based methods, the common approach is that a specific structure is imposed on a wave function, which is then optimized (often variationally) in order to approximate the ground state of a given system \cite{sakurai_modern_2010}. The success of any approach is then judged by how well the proposed structure of the wave function matches that of the exact ground state.

For integrable Richardson-Gaudin spin systems\cite{richardson_restricted_1963,richardson_exact_1964,gaudin_bethe_2014}, any eigenstate can be exactly written as \cite{links_completeness_2016}
\begin{equation}\label{bi:RGwf}
\ket{\psi_{RG}} = \prod_{\alpha=1}^N\left(\sum_{i=1}^L \frac{S^{\dagger}_i}{\epsilon_i - \lambda_{\alpha}}\right)\ket{\downarrow \dots \downarrow},
\end{equation}
where the different spins in the system are labelled $i=1, \dots, L$ and the spin operators constitute an $su(2)$ algebra (see Section \ref{sec:met}). The parameters $\epsilon_i, i=1 \dots L$ and $\lambda_{\alpha}, \alpha=1 \dots N$ have a clear physical interpretation within integrability, but can simply be thought of as arbitrary parameters for the time being. For this state to be an eigenstate of an integrable Hamiltonian, these variables are not independent and are coupled through the Bethe (or Richardson-Gaudin) equations
\begin{equation}\label{bi:RGeq}
1+\frac{g}{2} \sum_{i=1}^L \frac{1}{\epsilon_i-\lambda_{\alpha}}-g \sum_{\beta \neq \alpha}^N\frac{1}{\lambda_{\beta}-\lambda_{\alpha}}=0, \qquad \alpha=1 \dots N,
\end{equation}
where $g$ is an arbitrary parameter further tuning the correlations within the underlying integrable model. It is worth stressing that although all variables in this equation can be connected to the physics of an integrable system, it is not strictly necessary to interpret them as such. They can equally be treated as variational parameters, and we now propose to use this wave function as a variational ansatz. Given the Hamiltonian $\hat{H}$ of a strongly-correlated system, we wish to find the RG eigenstate that minimizes the energy (\ref{int:en}), resulting in a variational energy
\begin{equation}\label{bi:Evar}
E_{\textrm{Var.}} = \min E\left[\psi_{RG}\right].
\end{equation}
The scaling of this method is then set by the efficiency of the evaluation and minimization of the energy. In general, for arbitrary wave functions of the form (\ref{bi:RGwf}) (so-called \emph{off-shell} states), this scales exponentially with system size, 
but can be reduced to a polynomial complexity once the equations (\ref{bi:RGeq}) are satisfied (leading to \emph{on-shell} states), as shown in Section \ref{sec:met}. The exponential scaling explains why generalized spin states such as in Eq. (\ref{bi:RGwf}) have not attracted much consideration as a variational ansatz. For a generalized product state to be computationally tractable, it needs to be dressed with additional structure, which is here provided by integrability. It is worth noting that the projected BCS method can be reinterpreted as a special case of variational RG integrability, providing a connection between the variational wave function (\ref{bi:RGwf}) and the BCS mean-field wave function \cite{roman_large-n_2002}. Systems successfully described by mean-field theory, where the particles can be treated as non-interacting particles, also arise as a particular limit of the Bethe ansatz. In fact, a crucial feature of the wave function (\ref{bi:RGwf}) is that it exhibits a similar product structure as the Hartree-Fock wave function \cite{sakurai_modern_2010}. The variational method can thus already be expected to return accurate results for weakly-correlated systems.

The key question is then if the on-shell condition restricts the physics that can be captured by this ansatz. While integrable Hamiltonians are necessarily quite schematic, they have shown remarkable success in the description of general physical phenomena. Richardson's original solution to the (reduced) BCS Hamiltonian\cite{richardson_restricted_1963,richardson_exact_1964} already succeeded in qualitatively describing regular superconductivity\cite{bardeen_theory_1957,von_delft_spectroscopy_2001}, and only afterwards was it recognized that this Hamiltonian is integrable \cite{cambiaggio_integrability_1997}.

Furthermore, various efforts have shown how concepts from integrability may still prove useful when dealing with non-integrable systems. Form factors in integrable theories are exactly known, and can be used to build a perturbation theory for non-integrable models \cite{delfino_non-integrableqft_1996, controzzi_mass_2005,delfino_non-integrable_1998,pozsgay_characterization_2006,delfino_decay_2006,groha_spinon_2017}. Approximate scattering matrices for low-lying excited states of non-integrable systems have also been constructed from approximate (coordinate) Bethe ansatz techniques \cite{krauth_bethe_1991,kiwata_hirohito_bethe-ansatz_1994,okunishi_magnetization_1999,vanderstraeten_scattering_2015}. Despite these models being non-integrable, accurate results could still be obtained by applying techniques from integrability. Integrability-based methods have also been proposed in the description of time evolution governed by an integrable Hamiltonian plus a perturbation, both in the description of the initial behaviour \cite{brandino_relaxation_2013,van_den_berg_competing_2014} and the infinite-time behaviour \cite{lange_pumping_2017} of observables. Such problems have also been tackled using a numerical renormalization group expressed in the basis of eigenstates of the integrable model \cite{caux_constructing_2012}.

The majority of these results essentially build on the same idea as our proposed method - integrability can be used to describe the bulk of the correlations, on which corrections can be added. While using the same technical toolbox as these methods, our results are mainly similar in spirit to the use of perturbation theory for non-integrable system, where the important distinction is that the variational optimization guarantees a more accurate approximation of the ground state wave function than perturbation theory.

One final remark is that there exists no clear-cut definition of quantum integrability \cite{caux_remarks_2011}. Currently, the distinction between integrability and non-integrability is often made by numerically distinguishing statistical properties of the eigenvalue spectrum \cite{berry_level_1977,bohigas_characterization_1984}, where integrability-breaking leads to a crossover between two different behaviours \cite{poilblanc_poisson_1993,rabson_crossover_2004,relano_stringent_2004,santos_onset_2010,brandino_energy_2010}. Since this distinction occurs at the level of the total spectrum and not at the level of separate eigenstates, this suggests that, while it is not possible to approximate the total spectrum of a non-integrable model by an integrable model, it might still be possible to approximate the ground state of a non-integrable model by that of an integrable one, the main goal of the current work.

\section{Methods}
\label{sec:met}

In this work, we will make use of the theoretical and numerical framework underlying Richardson-Gaudin integrable models\cite{richardson_restricted_1963,richardson_exact_1964,gaudin_bethe_2014}. Their integrability and exact solvability have been derived in multiple ways \cite{amico_integrable_2001,amico_applications_2001,von_delft_algebraic_2002,zhou_superconducting_2002,links_algebraic_2003,ortiz_exactly-solvable_2005,skrypnyk_new_2005,skrypnyk_generalized_2007,skrypnyk_non-skew-symmetric_2009,skrypnyk_non-skew-symmetric_2009-1,skrypnyk_spin_2009,dukelsky_colloquium:_2004,dunning_exact_2010,links_completeness_2016}. We will give an overview of the main ingredients and refer the reader to Ref. [\onlinecite{ortiz_exactly-solvable_2005}] for a more extensive introduction to these systems and their applications.

\subsection{Richardson-Gaudin models}
The class of Richardson-Gaudin models\cite{richardson_restricted_1963,richardson_exact_1964,gaudin_bethe_2014} are based on the $su(2)$ algebra of (quasi-)spin operators \cite{talmi_simple_1993}. For systems describing the interactions between $L$ spins labelled $i=1 \dots L$, we first define a set of independent $su(2)$-algebras satisfying
\begin{align}
[S_i^0,S^{\dagger}_j]&=\delta_{ij}S^{\dagger}_i, \qquad  [S_i^0,S_j]=-\delta_{ij}S_i, \nonumber\\ {} [S^{\dagger}_i,S_j]&=2\delta_{ij}S_i^0.
\end{align}
One of the distinguishing characteristics of integrable systems is the existence of conserved charges, which are a set of mutually commuting operators in involution with the Hamiltonian. This means that each of these operators defines a quantity which is conserved under time evolution. The existence of a large amount of such conservation laws goes hand in hand with the existence of an exact solution, and is in fact one of the core aspects of quantum integrability \cite{caux_remarks_2011}. 

Richardson-Gaudin systems can then be defined through an explicit construction and parametrization of these conserved charges as
\begin{equation}\label{RG:com}
R_i = S_i^0 + g \sum_{j \neq i}^L \frac{1}{\epsilon_i-\epsilon_j}\left[\frac{1}{2}\left(S_i^{\dagger}S_j + S_i S_j^{\dagger}\right)+S_i^0 S_j^0\right].
\end{equation}
For any choice of the free variables $\vec{\epsilon}=\{\epsilon_1 \dots \epsilon_L\}$ these satisfy $[R_i,R_j]=0, \forall i,j=1 \dots L$. For our purpose, these variables will play the role of variational parameters in the eigenstates. An integrable Hamiltonian can be obtained by taking a linear combination of these operators as
\begin{equation}\label{RG:BAHam}
\hat{H} = \sum_{i=1}^L \eta_i R_i, \qquad \eta_i \in \mathbb{R}.
\end{equation}
Such a Hamiltonian is integrable, since it commutes with the conserved charges $R_i \  (i = 1 \dots L)$  by construction. The Hamiltonian and the conserved charges can then be simultaneously diagonalized by unnormalized Bethe ansatz eigenstates of the form
\begin{equation}
\ket{\vec{\epsilon},\vec{\lambda}} = \prod_{\alpha=1}^N S^{\dagger}(\lambda_{\alpha}) \ket{\downarrow \dots \downarrow},
\end{equation}
defined by a product of generalized raising operators
\begin{equation}
S^{\dagger}(\lambda) = \sum_{i=1}^L \frac{S_i^{\dagger}}{\epsilon_i-\lambda},
\end{equation}
where each generalized raising operator is fixed by a single parameter $\lambda \in \mathbb{C}$. This is known as a Bethe ansatz wave function, and is an eigenstate provided the variables $\vec{\lambda}=\{\lambda_1 \dots \lambda_N\}$ (also called rapidities) are coupled through the Richardson-Gaudin equations (\ref{bi:RGeq}). The generalized raising operators are fully determined by the relative position of $\lambda$ w.r.t. $\vec{\epsilon}$ in the complex plane, since the weight and phase of $S_i^{\dagger}$ follows from $(\epsilon_i-\lambda)^{-1}$. In this way, eigenstates can be determined by solving a set of coupled nonlinear equations scaling linearly in system size, which can be contrasted with the usual diagonalization of a Hamiltonian in an exponentially scaling Hilbert space. This is what is generally understood by exact solvability by Bethe ansatz. 

From the definition of the integrable Hamiltonian, we find that the most general Hamiltonian (containing maximally quadratic interactions) solvable by this method can be written as 
\begin{align}\label{RG:ham}
\hat{H} &= \sum_{i=1}^L \eta_i S_i^0 + \frac{g}{2} \sum_{i, j \neq i}^L \frac{\eta_i-\eta_{j}}{\epsilon_i-\epsilon_{j}} \left(\frac{1}{2}(S_i^{\dagger}S_j+S_i S_j^{\dagger})+S_i^0 S_j^0 \right) \nonumber \\
&=  \sum_{i=1}^L \eta_i S_i^0 + \frac{g}{2} \sum_{i, j \neq i}^L \frac{\eta_i-\eta_{j}}{\epsilon_i-\epsilon_{j}} \vec{S}_i \cdot \vec{S}_j,
\end{align}
for any set of variables $\vec{\epsilon}$ and $\vec{\eta}$. This Hamiltonian contains $2L$ free variables ($g$ can be absorbed in the definition of $\vec{\epsilon}$), and it is due to this freedom that we expect that Hamiltonians with similar interaction terms can be efficiently approximated within this approach. These Hamiltonians belong to the so-called rational (XXX) class of Richardson-Gaudin models, and can straighforwardly be extended towards hyperbolic (XXZ) models where the interaction between spins is anisotropic \cite{dukelsky_colloquium:_2004}. In the following, we will restrict ourselves to the rational models for clarity, but the proposed method can straightforwardly be extended towards these hyperbolic models.

\subsection{Calculating expectation values}
The building block for any variational method is the energy functional of a given wave function $E\left[\psi\right]$ (\ref{int:en}), which needs to be minimized with respect to the variational parameter defining the trial state $\ket{\psi} \equiv \ket{\vec{\epsilon}, \vec{\lambda}}$. The Bethe ansatz structure of the eigenstates allows for an efficient and relatively straightforward calculation of such expectation values, and can afterwards also be used to calculate observables from the obtained wave function. This expectation value is computationally tractable by making use of the overlap between an arbitrary (off-shell) Bethe state, and a state where the variables satisfy the Richardson-Gaudin equations (on-shell state). For off-shell Bethe states (\ref{bi:RGwf}) such expressions can only be evaluated through the use of extensive combinatorics, which cannot be evaluated in polynomial time, but the demand that the state is on-shell allows for simplifications \cite{amico_bethe_2012}. Once the variables satisfy the Richardson-Gaudin equations, inner products and expectations values can be expressed as determinants of matrices. Whole classes of such determinant expressions exist for this problem, following famous results by Slavnov\cite{slavnov_calculation_1989}, with the advantage that determinants can be efficiently evaluated numerically in polynomial time\cite{press_numerical_2007}. Suppose we have two states determined by the same set of variables $\vec{\epsilon}$ and different variables (rapidities) $\vec{v}=\{v_1 \dots v_N\}$ and $\vec{w}=\{w_1 \dots w_N\}$, where $\{v_1 \dots v_N\}$ satisfies the Richardson-Gaudin equations and $\{w_1 \dots w_N\}$ is arbitrary. Then the overlap between these two states is given by \cite{slavnov_calculation_1989}
\begin{align}
\braket{\vec{\epsilon},\vec{v}|\vec{\epsilon},\vec{w}} =& \frac{\prod_{b} \prod_{a \neq b} (v_a-w_b)}{\prod_{b<a} (w_b-w_a) \prod_{a<b}(v_b-v_a) } \nonumber \\
& \qquad \qquad \qquad \times \det S_N(\vec{v},\vec{w}),
\end{align}
with $S_N(\vec{v},\vec{w})$ an $N \times N$ matrix defined as
\begin{align}
S_N(\vec{v},\vec{w})_{ab} = \frac{v_b-w_b}{v_a-w_b}\bigg(&\sum_{i=1}^L \frac{1}{(v_a-\epsilon_i)(w_b-\epsilon_i)} \nonumber\\
&-2\sum_{c \neq a}^N \frac{1}{(v_a-v_c)(w_b-v_c)}  \bigg).
\end{align}
This is the well-known Slavnov determinant expression. Alternative determinant expressions can be found with a simpler structure, and it is possible to switch between determinant representations in order to control numerical stability \cite{claeys_inner_2017}.

From Slavnov's determinant, it immediately follows that the norm of an on-shell Bethe state can be calculated as the determinant of the Gaudin matrix
\begin{equation}
\braket{\vec{\epsilon},\vec{v}|\vec{\epsilon},\vec{v}} =\det{G_N(\vec{v})},
\end{equation}
with $G_N(\vec{v})$ an $N \times N$ matrix defined as 
\begin{equation}
 G_N(\vec{v})_{ab} =
  \begin{cases}
   \sum_{i=1}^L\frac{1}{(\epsilon_i-v_a)^2} -2 \sum_{c \neq a}^N \frac{1}{(v_c-v_a)^2}&\text{if}\ a=b \\
   \frac{2}{(v_a-v_b)^2}       & \text{if}\ a \neq b
  \end{cases}.
\end{equation}
Slavnov's determinant can then be used for the calculation of expectation values \cite{amico_exact_2002,links_algebraic_2003}, as illustrated in Appendix \ref{app:corr}. Here, the key feature is that the action of any Hamiltonian on an on-shell Bethe state can be written as a (polynomially large) sum of off-shell Bethe states, so expectation values can always be written as a polynomial summation of Slavnov determinants. This, combined with the determinant for the normalizations, allows the variational energy (\ref{bi:Evar}) to be evaluated in a polynomial time for on-shell states. We refer to Appendix \ref{app:corr} for an analysis of the computational scaling.

\subsection{Numerics}
The framework of integrability reduces the problem of finding eigenstates of a Hamiltonian to solving a set of nonlinear equations (\ref{bi:RGeq}), and numerical methods have been tailored to this specific problem \cite{rombouts_solving_2004,rombouts_quantum_2010,guan_heine-stieltjes_2012,pan_heine-stieltjes_2013,guan_numerical_2014,qi_exact_2015}. In practice, the Richardson-Gaudin equations (\ref{bi:RGeq}) are rarely solved directly because they exhibit singular behaviour. 

A common approach, known as the eigenvalue-based method, maps the Richardson-Gaudin equations (\ref{bi:RGeq}) to an equivalent set of equations for the variables\cite{babelon_bethe_2007,faribault_gaudin_2011,faribault_determinant_2012,el_araby_bethe_2012,tschirhart_algebraic_2014,claeys_eigenvalue-based_2015,faribault_determinant_2016,claeys_eigenvalue-based-bosonic_2015,claeys_read-green_2016,links_completeness_2016,faribault_common_2017} 
\begin{equation}
\Lambda_i = \sum_{\alpha=1}^N \frac{1}{\epsilon_i-\lambda_{\alpha}}, 
\end{equation}
which satisfy the set of quadratic equations
\begin{equation}\label{met:EVBeq}
\Lambda_i^2 = -\frac{2}{g} \Lambda_i+ \sum_{j \neq i}^L \frac{\Lambda_i-\Lambda_j}{\epsilon_i-\epsilon_j}, \qquad i=1, \dots , L.
\end{equation}
Because these equations are quadratic, they are more stable than the original Richardson-Gaudin equations, which suffer from numerical singularities at the so-called `singular points'\cite{richardson_numerical_1966,dominguez_solving_2006,de_baerdemacker_richardson-gaudin_2012}. While the number of equations that needs to be solved has increased ($L$ compared to $N$), these remain within the same order of magnitude and the increase in numerical stability more than makes up for this. Eq. (\ref{met:EVBeq}) can then be solved using iterative methods such as the Newton-Raphson method \cite{press_numerical_2007}. Once these variables have been obtained, the rapidities $\vec{\lambda}$ still need to be determined for the calculation of expectation values (see Appendix \ref{app:corr}). One way this can be realized is by defining a polynomial with the rapidities as roots
\begin{equation}
P(z) = \prod_{a=1}^N (z-v_a),
\end{equation}
which satisfies the ordinary differential equation (ODE) \cite{dorey_ode/im_2007,el_araby_bethe_2012,links_completeness_2016}
\begin{equation}\label{met:ODE}
P''(z) + F(z) P'(z)-G(z)P(z)=0,
\end{equation}
with 
\begin{align}
F(z) = \frac{2}{g}+\sum_{i=1}^L \frac{1}{\epsilon_i-z}, \qquad
G(z) = \sum_{i=1}^L \frac{\Lambda_i}{\epsilon_i-z}.
\end{align}
Once the variables $\Lambda_i$ have been obtained, the differential equation (\ref{met:ODE}) is fixed and efficient algorithms have been developed to find the roots of this polynomial, extracting the rapidities \cite{el_araby_bethe_2012}. Solving for an eigenstate thus consists of a two-part process: first a set of quadratic equations are solved for the variables $\Lambda_i$, after which the rapidities are obtained by using the BA/ODE correspondence. This method provides accurate results for models with up to a few hundred spin levels.

\subsection{Optimizing the wave function}
So, for a given Hamiltonian $\hat{H}=\hat{H}_{int}+\hat{V}$, with $\hat{H}_{int}$ an integrable (Richardson-Gaudin) Hamiltonian, and $\hat{V}$ containing additional interactions breaking the integrability, we wish to minimize
\begin{equation}
E\left[\psi_{RG}\right] = \frac{{\braket{\psi_{RG}|\hat{H}|\psi_{RG}}}}{\braket{\psi_{RG}|\psi_{RG}}},
\end{equation}
where $\ket{\psi_{RG}} \equiv \ket{\vec{\epsilon}, \vec{\lambda}}$, with $\vec{\epsilon}=\{\epsilon_1 \dots \epsilon_L\}$ and $\vec{\lambda}=\{\lambda_1 \dots \lambda_N\}$ coupled through the Richardson-Gaudin equations. It is important to note that the variables in the wave function are independent from those in the integrable Hamiltonian, since the former are the degrees of freedom over which we optimize, while the latter are a characteristic of the unperturbed system. Obviously, in the limit of a vanishing perturbation $\hat{V}=0$ the Hamiltonian $\hat{H}$ becomes integrable, and the variational optimization should return the variables in the Hamiltonian as variational parameters, since this wave function is then the exact ground state of the integrable Hamiltonian.

While these states explicitly depend on $L+N$ variables $\vec{\epsilon}$ and $\vec{\lambda}$, the demand that these states are on-shell (\ref{bi:RGeq}) leaves us with $L$ degrees of freedom over which to optimize, which we choose as $\vec{\epsilon}$, and we can simply denote $E[\psi_{RG}]=E[\vec{\epsilon}]$, with the implicit assumption that all rapidities $\vec{\lambda}$ uniquely follow through the Richardson-Gaudin equations, resulting in a manifold of states only determined by the variables $\vec{\epsilon}$. However, we have an additional discrete degree of freedom - the choice of eigenstate. Each eigenstate of an integrable Hamiltonian defined by a set of variables $\vec{\epsilon}$ can be written as (\ref{bi:RGwf}), so we need to somehow specify what state we wish to target. This degree of freedom will initially be disregarded, and we will restrict ourselves to the state that is adiabatically connected to the ground state of the integrable Hamiltonian in the limit of a vanishing perturbation. For small perturbations, it is expected that this state will be the most relevant. Later, it will be shown that this choice is not guaranteed to be optimal for large perturbations, and the excited states will prove to be important. 

We choose to optimize over the variables $\vec{\epsilon}$ using a gradient descent method \cite{press_numerical_2007}. The necessary ingredient for this algorithm is the gradient of the function to be minimized, which is here calculated by a finite difference estimation using a numerically small step $\Delta \epsilon$ for a two-point estimation to obtain
\begin{equation}
\left(\vec{\nabla} E\left[\vec{\epsilon}\right] \right)_i \approx \frac{E\left[\vec{\epsilon}+\Delta \epsilon \cdot  \vec{1}_i \right] -E\left[\vec{\epsilon}-\Delta \epsilon \cdot \vec{1}_i\right] }{2 \Delta \epsilon} .
\end{equation}
It should be noted that a change in one of the variables $\vec{\epsilon}$ also implies a resulting change in all rapidities $\vec{\lambda}$, since we demand the wave function to be on-shell at each step of the calculation. Often, this can be done using a straightforward Newton-Raphson approach, but care should be taken when multiple rapidities are close together. In our approach, we optimize the eigenvalue-based variables $\Lambda_i$ using a Newton-Raphson approach, and afterwards extract the updated variables using the BA/ODE correspondence starting from the previous rapidities.

So, our approach can be summarized in Algorithms (\ref{alg:sd}) and (\ref{alg:upd}). While the gradient descent method is a straightforward one, care should again be taken with the implicit dependence of the rapidities $\vec{\lambda}$ on the variables $\vec{\epsilon}$. This is illustrated in Algorithm \ref{alg:upd}, exploiting previous numerical work on Richardson-Gaudin models. The procedure outlined here takes care to avoid the singularities arising in the Richardson-Gaudin equations. 

\begin{algorithm}[H]
\caption{Variational optimization of an on-shell Bethe ansatz state.\label{alg:sd}}
\begin{algorithmic}
\State Define $\hat{H}=\hat{H}_{int}+\hat{V}$
\State Define $\vec{\epsilon}_0$ \Comment{Follows from $\hat{H}_{int}$}
\State Define $\vec{\lambda}_0$ \Comment{Solve RG eq. given $\vec{\epsilon}_0$}
\State $\vec{\epsilon},\vec{\lambda} \gets \vec{\epsilon}_0, \vec{\lambda}_0$
\State $\Delta E \gets 0$
\While {$\Delta E < 0$} \Comment{Update state while energy decreases.}
\State Calculate $E[\vec{\epsilon}]$ and  $\nabla E[\vec{\epsilon}]$ \Comment{Update $\vec{\lambda}$ for gradient.}
\State $\mu \gets 0$
\While {$E[\vec{\epsilon}-\mu \vec{\nabla} E[\vec{\epsilon}] ] <  E[\vec{\epsilon}]$}
    \State Increase $\mu$
    \State Calculate $E[\vec{\epsilon}-\mu \vec{\nabla} E[\vec{\epsilon}] ]$ \Comment{See Algorithm \ref{alg:upd}.}
\EndWhile
\State $\Delta E \gets  E[\vec{\epsilon}-\mu \nabla E[\vec{\epsilon}] ] - E[\vec{\epsilon}]$
\State $\vec{\epsilon} \gets \vec{\epsilon}-\mu \nabla E[\vec{\epsilon}]$ \Comment{Update state.}
\EndWhile
\State Optimized energy $E[\vec{\epsilon}]$
\State Optimized state $\ket{\vec{\epsilon},\vec{\lambda}}$
\end{algorithmic}
\end{algorithm}

\begin{algorithm}[H]
\caption{Update energy $E[\vec{\epsilon}]$ to $E[\vec{\epsilon}+\vec{\delta}]$ for an on-shell Bethe ansatz state. \label{alg:upd}}
\begin{algorithmic}
\State Define $\hat{H}$
\State Define $\vec{\epsilon}$, $\vec{\lambda}$, $\vec{\Lambda}$ \Comment{Known from previous calc.}
\State $\vec{\epsilon} \gets \vec{\epsilon}+\vec{\delta}$
\State $\vec{\Lambda}' \gets $Update $\vec{\Lambda}$ from $\vec{\epsilon}+\vec{\delta}$ \Comment{Substituted quadr. eq.}
\State $\vec{\lambda}' \gets $ Update $\vec{\lambda}$ from $\vec{\Lambda}$ \Comment{BA/ODE}
\State $E[\vec{\epsilon}+\vec{\delta}] \gets \braket{\vec{\epsilon}+\vec{\delta}, \vec{\lambda'}|\hat{H}|\vec{\epsilon}+\vec{\delta}, \vec{\lambda'}}/\braket{\vec{\epsilon}+\vec{\delta}, \vec{\lambda'}|\vec{\epsilon}+\vec{\delta}, \vec{\lambda'}}$
\end{algorithmic}
\end{algorithm}
\section{Results}
\label{sec:res}
In this section, we will apply the proposed algorithm to the two predominant classes of Richardson-Gaudin integrable models: the central spin model \cite{gaudin_bethe_2014} and the reduced BCS (Richardson) Hamiltonian \cite{richardson_restricted_1963,richardson_exact_1964}. 

\subsection{Perturbing the central spin model}
The results will first be illustrated on the central spin model with perturbations restricted to operators acting on one or two spins. The central spin Hamiltonian is given by
\begin{equation}\label{res:csHam}
\hat{H}_{cs} = B S_1^z + g \sum_{i \neq 1}^L \frac{\vec{S}_1 \cdot \vec{S}_i}{\epsilon_{0,1}-\epsilon_{0,i}} ,
\end{equation}
describing the interaction of a single spin (where we have identified $S_1^z \equiv S_1^0$), on which a magnetic field $B$ is applied, with a bath of surrounding spins. This model is often studied in the context of NV centers or semiconductor quantum dots \cite{coish_hyperfine_2004,faribault_spin_2013,faribault_integrability-based_2013,van_den_berg_competing_2014}. The Hamiltonian (\ref{res:csHam}) equals one of the conserved charges (\ref{RG:com}), and is thus integrable for any choice of the interaction modulated by $\epsilon_{0,i}$, which has been written in this way in order to make this connection explicit. 

In this model, the bath spins do not interact among themselves and do not experience the magnetic field applied to the central spin. However, such interactions may be added in a perturbative way by introducing terms of the form $S_i^0$ and $\vec{S}_i \cdot \vec{S}_j$ in the Hamiltonian. The basic physics in this model can be easily understood -- $B$ determines the orientation of the central spin $\braket{S_1^z}$, either parallel or anti-parallel to the quantization axis, while the signs of $g/(\epsilon_{0,1}-\epsilon_{0,i})$ determine the relative orientation of the bath spin $\vec{S}_i$ with the central spin $\braket{\vec{S}_1 \cdot \vec{S}_i}$. 

In the following, we considered system sizes $L=12$ for which exact diagonalization methods can be used as benchmark and parametrize the Hamiltonian with $\epsilon_{0,i} = L- i$ as a picket-fence model \cite{hirsch_fully_2002}. The strength of the interaction is fixed by setting $B=1$ and $g=-2$, intermediate between strong- and weak-coupling \cite{de_baerdemacker_richardson-gaudin_2012}. For this choice of parametrization, the central spin and all surrounding spins tend to align, while also being restricted by conservation of spin projection $S^z = \sum_i S_i^z$. In the following, we always choose $S^z=0$ (or $L=2N$), since this is the sector where the dimension of the Hilbert space is maximal.

\begin{center}
\begin{figure}
\includegraphics[width=\columnwidth]{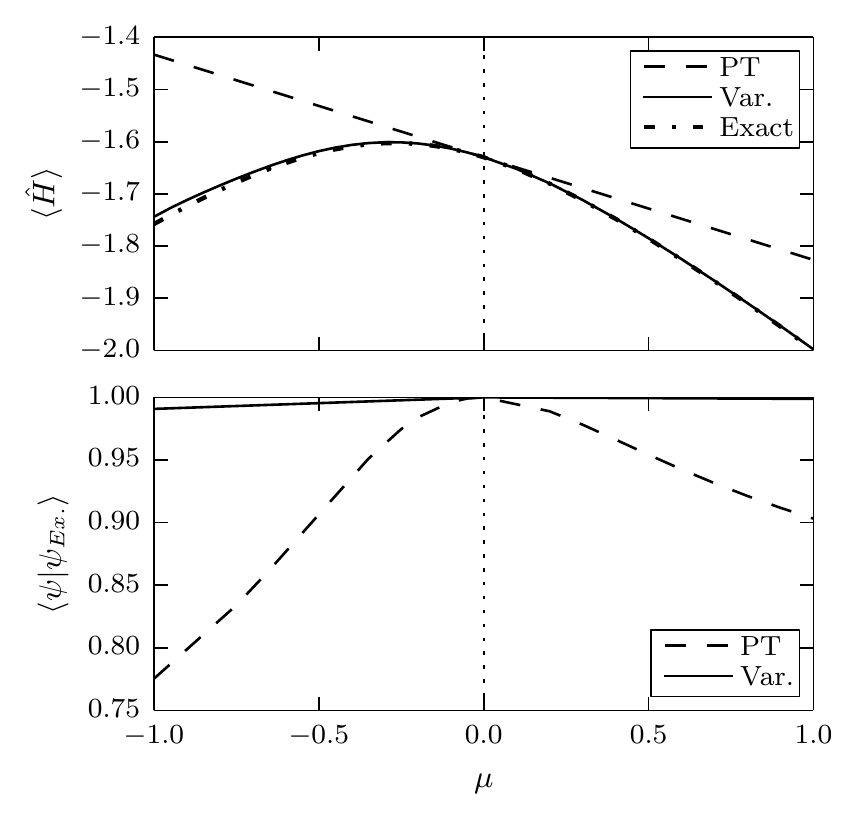}
\caption{Results for the central spin model with perturbation $\mu S_i^0$. \textbf{Top}: Variational energy (Var.), exact ground state energy (Exact), and first-order perturbation theory (PT) energy for different values of the perturbation strength. \textbf{Bottom}: Overlap of the exact ground state with the variational ground state and the ground state of the unperturbed model.
 \label{fig:cs:ener_ov_Ji}}
\end{figure}
\end{center}

\textbf{Single-spin perturbation.} Firstly, we perform calculations for a Hamiltonian 
\begin{equation}\label{res:HamJi}
\hat{H} = \hat{H}_{cs}+\mu S_i^0,
\end{equation}
applying a magnetic field of size $\mu$ to one of the spins in the bath (here labelled $i$). Such a model has previously also been investigated in the context of integrability-breaking \cite{erbe_different_2010,schliemann_spins_2010}. Here, we calculate the variational energy and compare with the ground-state energy obtained by exact diagonalization. In Figure \ref{fig:cs:ener_ov_Ji}, we plot the variational energy (Var.), the exact ground-state energy (Exact), and the energy obtained by first-order perturbation theory (PT) for varying perturbation strengths $\mu$, with $i=2$ chosen to maximize the deviation from the integrable model, since the central spin interacts most strongly with this bath spin. Since the ground-state energy deviation is intimately connected to the overlap between the approximate ground state and the exact ground state, this is also given in Figure \ref{fig:cs:ener_ov_Ji}. As the error in the energy is generally quadratic in the error in the overlap, the latter can be seen as a more sensitive measure for the accuracy of the proposed method.

Labelling the parameters of the unperturbed integrable model as $\vec{\epsilon}_0$, the relevant energies can be contrasted as
\begin{align}
E_{\textrm{Var.}} = \min_{\vec{\epsilon}} E\left[\vec{\epsilon}\right], \qquad E_{\textrm{PT}} = E\left[\vec{\epsilon}_0\right],
\end{align}
making clear why the variational method provides a guaranteed improvement on first-order perturbation theory. In the chosen model, perturbation theory is guaranteed to provide a good approximation to the exact ground state energy only when $|\mu \braket{S_i^0}| \ll\Delta E$, with $\Delta E$ the energy difference between the ground state and the first excited state. In the following, this roughly corresponds to $|\mu| \ll 1$, which we will consider to be a small perturbation.
The overlaps given are those between the variationally obtained wave function (Var.) and the exact ground state, together with the overlap between the ground state of the unperturbed model and the exact ground state (PT). The variational wave function is able to accurately model the ground state for a wide range of the perturbation strength, even going up to the limit where the size of the perturbation interaction equals that of the unperturbed central spin interaction ($|\mu|=1$), providing a substantial improvement over first-order perturbation theory. Here, the variational optimization plays a crucial role, as can be seen by comparing the overlap of the exact ground state with the ground state of the unperturbed Hamiltonian to the overlap with the variationally optimized wave function, which is improved by several orders of magnitude (from an overlap of 0.7754 to 0.9908 for $\mu=-1$). However, since the perturbation here only acts on a single spin site, it is not expected that this will fundamentally influence the correlations in the model, and more intrusive perturbations may be more physical.

\begin{figure*}[tb!]                 
 \begin{center}
 \includegraphics{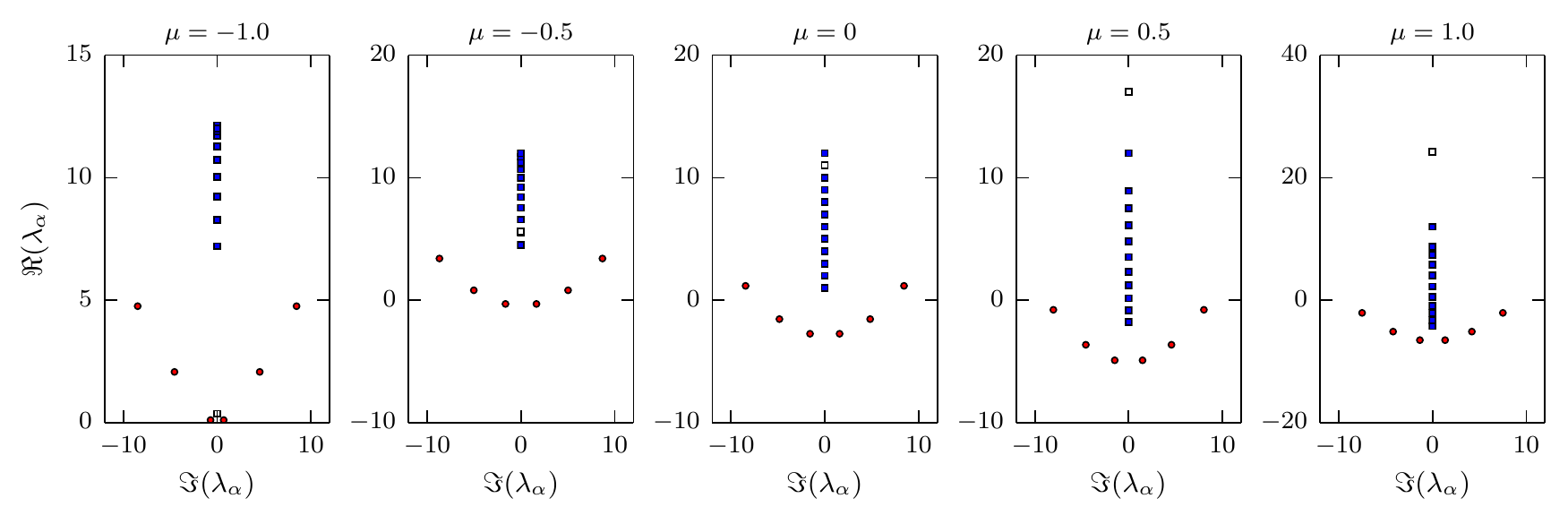} 
 \caption{Variational parameters for the Hamiltonian (\ref{res:HamJi}). Position of $\vec{\epsilon}$ (squares) and $\vec{\lambda}$ (dots) for the variationally optimized wave function in the complex plane at different values of the perturbation strength. The white square denotes the variable $\epsilon_i$ associated with the level on which the perturbation is applied.
\label{fig:rap_Ji}}
 \end{center}
\end{figure*}

Some more insight in the role of the optimization and the structure of the wave function can be obtained by considering the evolution of the variables $\vec{\epsilon}$ and $\vec{\lambda}$ in the wave function. These are given in Figure \ref{fig:rap_Ji} for different values of the perturbation strength. The variables $\vec{\epsilon}$ are restricted to be real, while the rapidities $\vec{\lambda}$ are either real or arise as complex conjugate pairs. The single-spin character of the perturbation is clear from these figures. Only the variable $\epsilon_i$ ($i=2$), associated to the perturbed level, is significantly sensitive to the perturbation, whereas all other variables are largely unaffected. While the on-shell condition still connects both sets of variables, it can be seen that the variables $\vec{\lambda}$ are quite robust against perturbations. This also motivates the use of $\vec{\epsilon}$ as variational parameters.

\begin{center}
\begin{figure}
\includegraphics[width=\columnwidth]{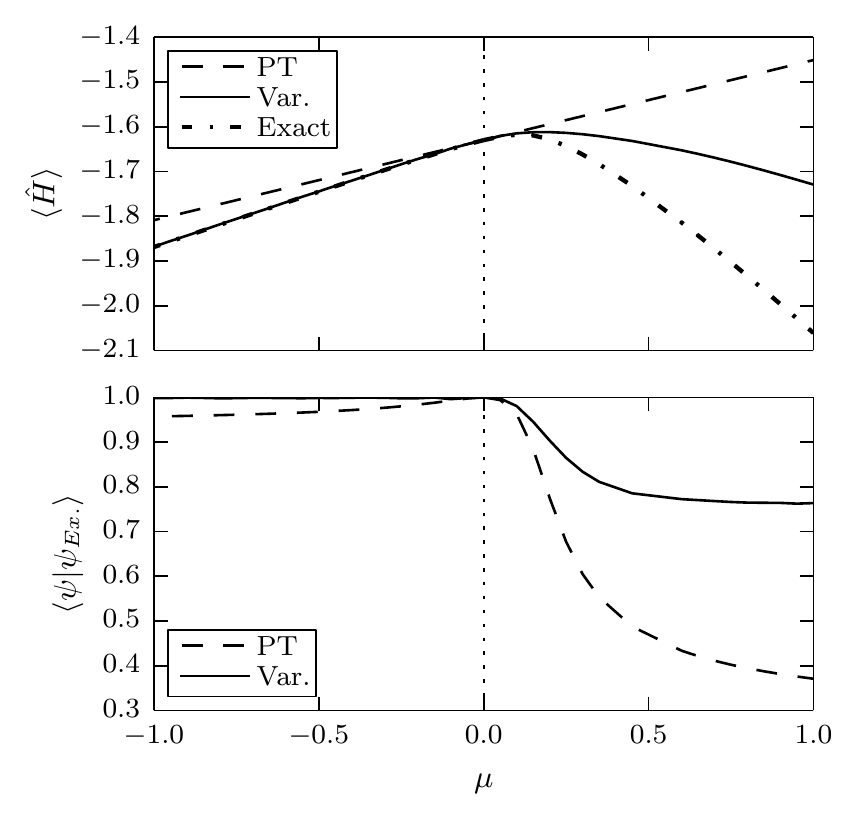}
\caption{Results for the central spin model with perturbation $\mu \vec{S}_i \cdot \vec{S}_j$, with $i,j=2,L-1$. \textbf{Top}: Variational energy, exact ground state energy, and first-order perturbation theory energy for different values of the perturbation strength. \textbf{Bottom}: Overlap of the exact ground state with the variational ground state and the ground state of the unperturbed model. \label{fig:ener_ov_Jij}}
\end{figure}
\end{center}

\textbf{Double-spin perturbation.}  Secondly, this method is applied to a non-integrable Hamiltonian
\begin{equation}\label{res:HamJij}
\hat{H} = \hat{H}_{cs} + \mu \vec{S}_i \cdot \vec{S}_j,
\end{equation}
where $\mu$ again determines the perturbation strength, and repeat the same calculations, where we choose bath spins $i,j=2,L-1$ for similar reasons as before. The results for the energy and overlap are given in Figure \ref{fig:ener_ov_Jij}.
It can be seen that the variational method still provides an accurate description for negative $\mu$, but interestingly fails to model the behaviour of the wave function for large positive $\mu$. The method holds in the limit where we can interpret the additional term as a perturbation ($|\mu \braket{\vec{S}_i \cdot \vec{S}_j} | \ll \Delta E$), but moving away from this limit the method quickly breaks down.

The reason for this can be inferred from perturbation theory for the two different regimes (positive and negative $\mu$). In the ground state of the unperturbed model $\braket{\vec{S}_i \cdot \vec{S}_j} > 0$, since all spins tend to align. So, the perturbation will lower the ground state energy if $\mu<0$ and increase the energy if $\mu>0$. In the former case, the perturbation does not qualitatively change the physics in the model, whereas the latter introduces a counteracting interaction, lowering the energy if the two spins are anti-parallel. For larger $\mu$ ($\mu 
\gtrsim 0.2$),  the energy then again lowers, pointing to a change in qualitative character of the ground state. The sudden drop in overlap with the exact ground state in Figure \ref{fig:ener_ov_Jij} then hints at an avoided crossing between the ground state and an excited state for increasing $\mu$, where if $\mu$ is increased the ground state would resemble an excited state of the original system rather than the ground state. The variational optimization is still capable of increasing the overlap by more than a factor 2, but is ultimately unable to obtain an accurate description for large $\mu$. This can be understood since, while the perturbation increases the energy of the unperturbed ground state, it simultaneously lowers the energy of selected excited states of the unperturbed model.

For relatively simple perturbations, the relevant excited state can be gathered from the limit $|\mu| \to \infty$, where the perturbation becomes dominant, and we can variationally optimize the state which is adiabatically connected to this excited state in the limit $\mu \to 0$.
\begin{center}
\begin{figure}
\includegraphics[width=\columnwidth]{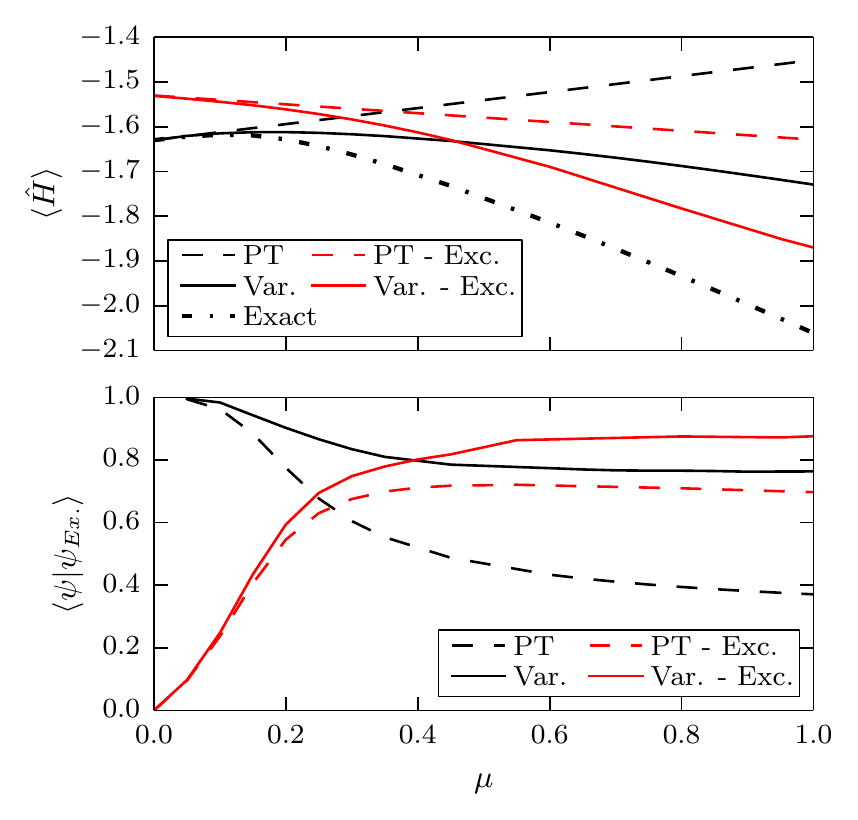}
\caption{Results for the central spin model with perturbation $\lambda \vec{S}_i \cdot \vec{S}_j$. \textbf{Top}: Variational energy, exact ground state energy, and first-order perturbation theory energy starting from the ground and excited state of the integrable model for different values of the perturbation strength. \textbf{Bottom}: Overlap of the exact ground state with the variational ground state and the ground state of the unperturbed model starting from both the ground and excited state of the integrable model. The perturbation has been applied to the spin with the strongest interaction with the central spin in order to maximize the effect of the perturbation.\label{fig:ener_ov_Jij_exc}}
\end{figure}
\end{center}
For positive $\mu$, the results for a variational optimization starting from both the ground state and this excited state are presented in Figure \ref{fig:ener_ov_Jij_exc}. At small $\mu$, the unperturbed ground state is the energetically favourable one, while for increasing perturbation strength the energy of the unperturbed excited state drops below that of the unperturbed ground state. Such crossings are observed both in perturbation theory and in the variational method, albeit occurring for smaller values of the perturbation in the variational method. This behaviour can also be observed from the overlaps, where a similar crossing occurs in the same region. The variational optimization again plays an important role in lowering the energy and increasing the overlap, both for the variational state obtained from the unperturbed ground- and excited state, resulting in an improved approximation to the ground state. Note that, while this results in a much improved description, there is still a part of the wave function that cannot be captured by the variational method, and for which we would either need to apply perturbation theory on the optimized wave function, or use a multi-reference approach with multiple Bethe ansatz wave functions in the variational optimization. However, the technicality of these approaches exceed the range of the current article.

\begin{figure*}[tb!]                 
 \begin{center}
 \includegraphics{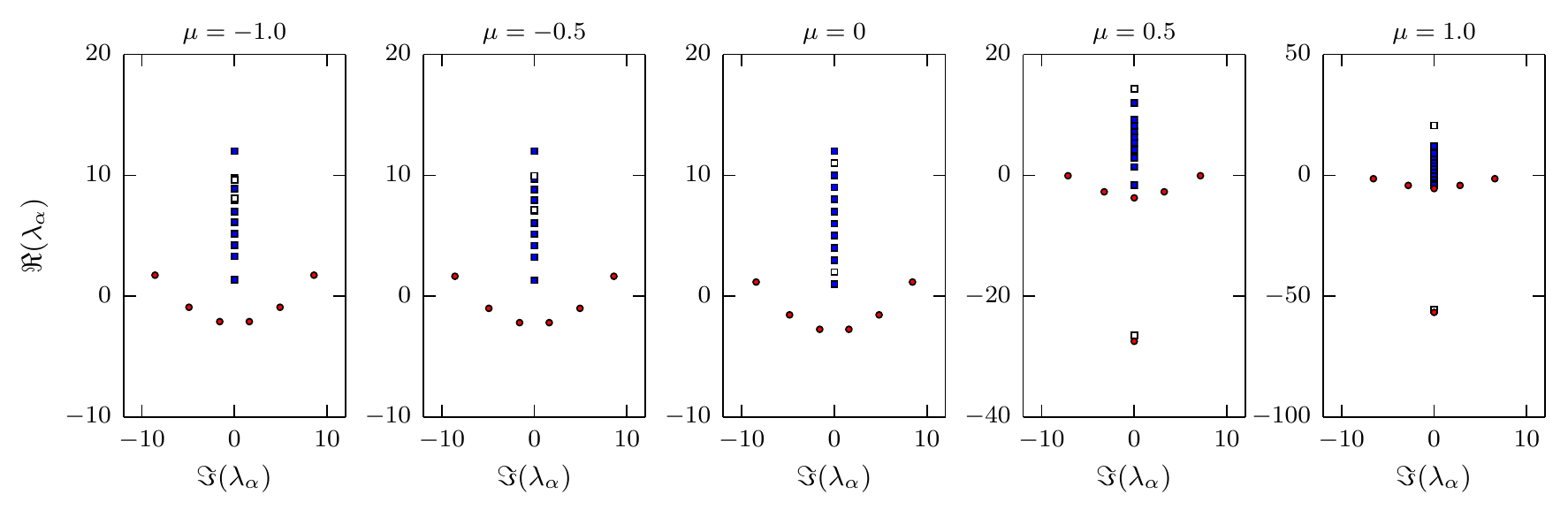} 
 \caption{Variational parameters for the Hamiltonian (\ref{res:HamJij}). Position of $\vec{\epsilon}$ (squares) and $\vec{\lambda}$ (dots) for the variationally optimized wave function in the complex plane at different values of the perturbation strength. The white squares denote the variables $\epsilon_i, \epsilon_j$ associated with the levels on which the perturbation is applied. \label{fig:rap_Jij_gs}}
 \end{center}
\end{figure*}

\begin{center}
\begin{figure}
\includegraphics[width=\columnwidth]{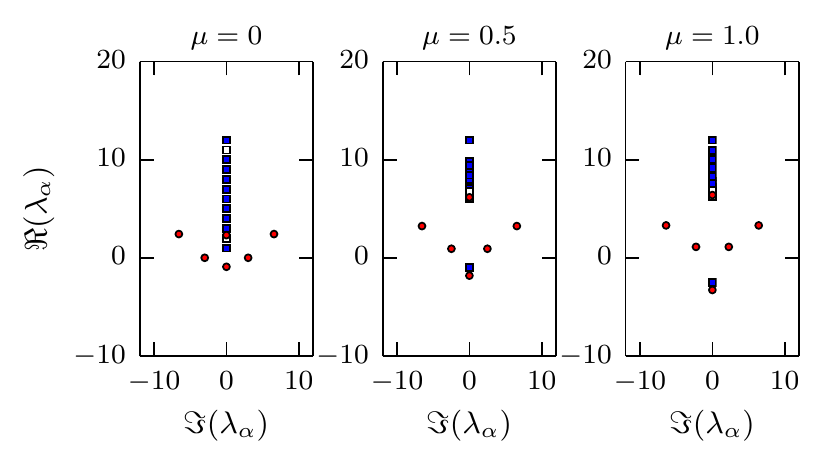}
\caption{Results for the central spin model with perturbation $\lambda \vec{S}_i \cdot \vec{S}_j$. Position of $\vec{\epsilon}$ (squares) and $\vec{\lambda}$ (dots) for the variationally optimized wave function starting from an excited state in the complex plane at different values of the perturbation strength. The white squares denote the variables $\epsilon_i, \epsilon_j$ associated with the levels on which the perturbation is applied. \label{fig:rap_Jij_exc}}
\end{figure}
\end{center}

\begin{figure*}[htb!]                 
 \begin{center}
 \includegraphics{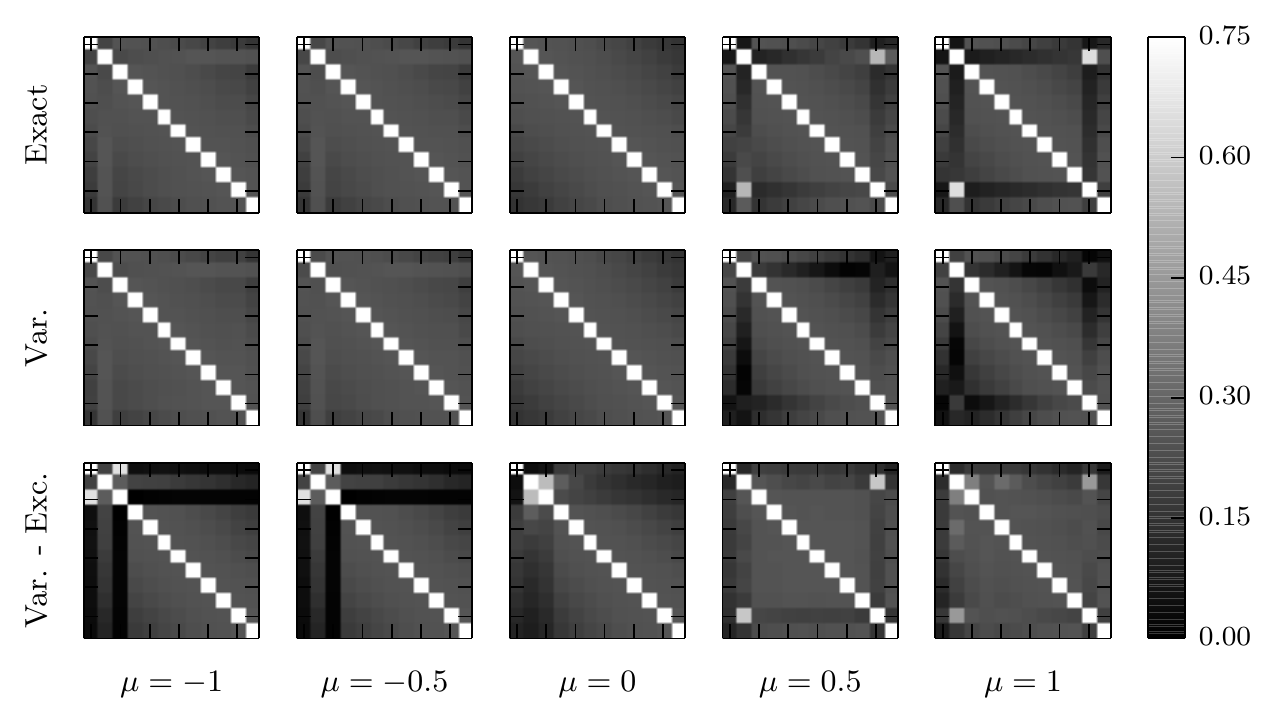} 
 \caption{Absolute value of the expectations values $\braket{\vec{S_i} \cdot \vec{S}_j}$ for the central spin model with perturbation $\mu \vec{S}_i \cdot \vec{S}_j$, with $i,j=2,L-1$. The expectation values are taken w.r.t. the exact ground state and the variational state obtained by starting from both the ground state and excited state of the unperturbed model. \label{fig:exp_Jij}}
 \end{center}
\end{figure*}

\begin{figure*}[htb!]              
 \begin{center}
 \includegraphics{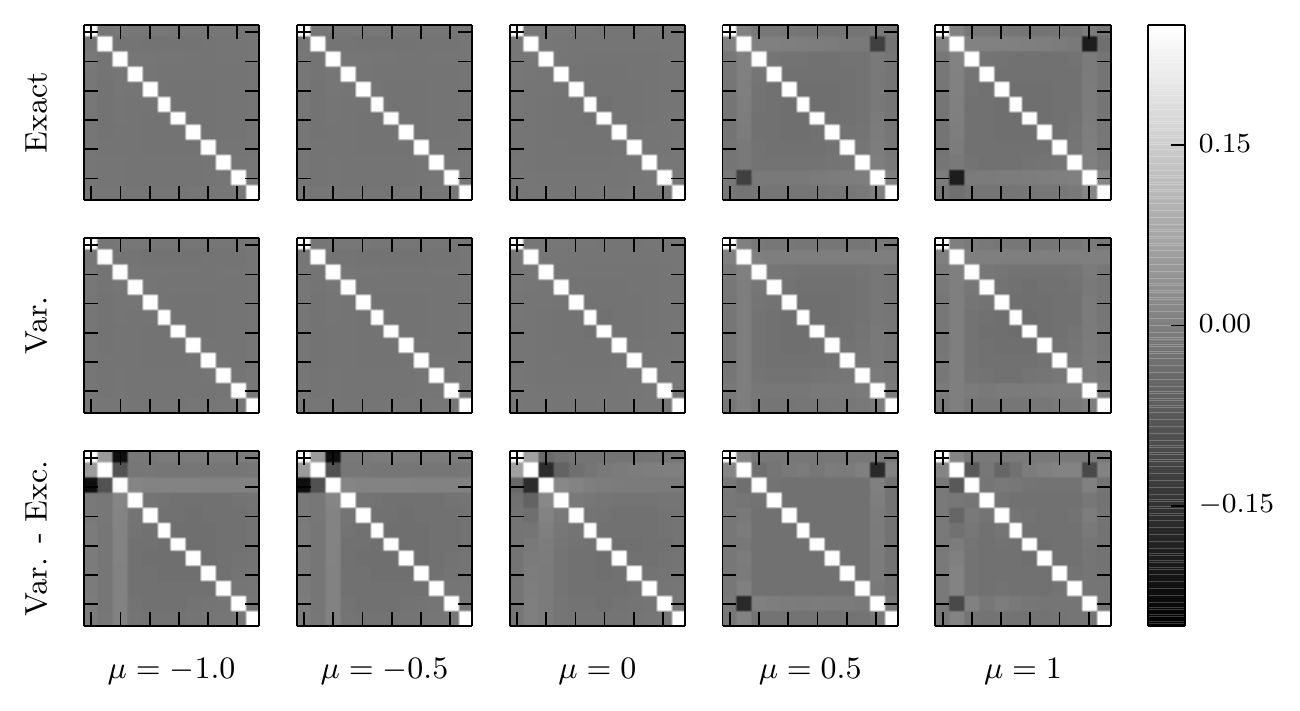} 
 \caption{Correlation coefficients $\braket{S_i^0 S_j^0}-\braket{S_i^0}\braket{S_j^0}$ for the central spin model with perturbation $\mu \vec{S}_i \cdot \vec{S}_j$, with $i,j=2,L-1$. These correlation coefficients are calculated for the exact ground state and the variational state obtained by starting from both the ground state and excited state of the unperturbed model.  \label{fig:corr_Jij}}
 \end{center}
\end{figure*}

The structure of the optimized variables $\vec{\epsilon}$ and $\vec{\lambda}$ can again be compared (Figures \ref{fig:rap_Jij_gs} and \ref{fig:rap_Jij_exc}). The two-spin character of the interaction is clearly visible, where the optimization is mainly sensitive to the two variables $\epsilon_i, \epsilon_j$ $(i=2, j=11)$ in the region where the optimization performs well. When the optimization fails to provide an accurate wave function, the rapidities exhibit a qualitative change (complex conjugate variables become real) and quickly increase in absolute value, pointing out that they are qualitatively wrong. Starting from the excited state in the unperturbed model, it is observed that the rapidities already have the correct structure, and remain bounded during the optimization.
Some more physical insight can be gathered from expectation values and correlation coefficients calculated from both wave functions. In Figures \ref{fig:exp_Jij} and \ref{fig:corr_Jij}, we present the expectation values $\braket{\vec{S_i} \cdot \vec{S_j}}, \forall i,j$, motivated by the choice of perturbation interactions, and the unconnected correlation coefficients $\sigma_{ij} = \braket{S_i^0S_j^0}-\braket{S_i^0}\braket{S_j^0}, \qquad \forall i,j$ for both exact and variational wave functions at different values of $\mu$. It is clear that the correlations within the wave function only change slightly for negative $\mu$, and as such the wave function is able to easily adapt to the perturbation. Comparing the exact and the variational ground state for positive $\mu$, it is notable that the correlations between the two spins affected by the interactions have not been captured by the variational ground-state wave function. Comparing this with the results from the variational excited wave function, it can be seen that the missing correlations are reintroduced there, as was expected. 
For low perturbation strengths, the variational wave function is able to adapt to the correlation structure of the exact ground state  through the optimization. The change in correlation coefficients also points towards the failure of perturbation theory. In the region $\mu 
\gtrsim 0.2$, the unconnected correlation coefficients from the approximate wave function for the levels on which we apply the perturbation vanish, and this level effectively decouples from the many-body system. In the exact wave function, this decoupling does not occur and instead these coefficients change sign. From this, it can be concluded that the wave function can adapt to the perturbation for as long as the general structure of the correlations does not change. By starting the variational optimization from the excited state, the correct structure is again recovered, as can be seen in the bottom row of Figures \ref{fig:exp_Jij} and \ref{fig:corr_Jij}.

This can now also be compared to the expected range of applicability of perturbation theory. For the given Hamiltonians $\hat{H}=\hat{H}_{cs}+\mu \hat{V}$, perturbation theory starting from the integrable $\mu=0$ limit can be expected to provide accurate results only if $|\mu \braket{\hat{V}}| \ll \Delta E$, in the regime where the additional term can be considered a small perturbation on the integrable model. The variational optimization starting from the ground state results in a relatively accurate approximation for a larger range of $\mu$, even when the additional term can no longer be considered to be a small perturbation, provided there occur no avoided crossings between the ground- and excited states of the integrable Hamiltonian in the spectrum of the non-integrable Hamiltonian when the perturbation strength $\mu$ is adiabatically increased from $0$ to the given value. Because these Hamiltonians are non-integrable these are expected to be avoided crossings, but this reasoning should also hold for allowed level crossings.

We have checked that the same behaviour is observed when introducing more involved perturbations, where small perturbations can be accurately described starting from the ground state of the integrable Hamiltonians, and for larger perturbations variational optimization starting from an excited state is necessary in order to obtain the optimal approximative state. However, at present it is not always clear which excited state should be chosen for arbitrary perturbations. In practice, this problem could be circumvented using a stochastic approach, since it was found that several excited states can lead to the same variationally optimized state. In practice, all relevant excited states for considered perturbations were obtained as so-called 1p-1h or 2p-2h excitations of the ground state \cite{dukelsky_colloquium:_2004}.

\subsection{Perturbing the Richardson model}
\begin{center}
\begin{figure}
\includegraphics[width=\columnwidth]{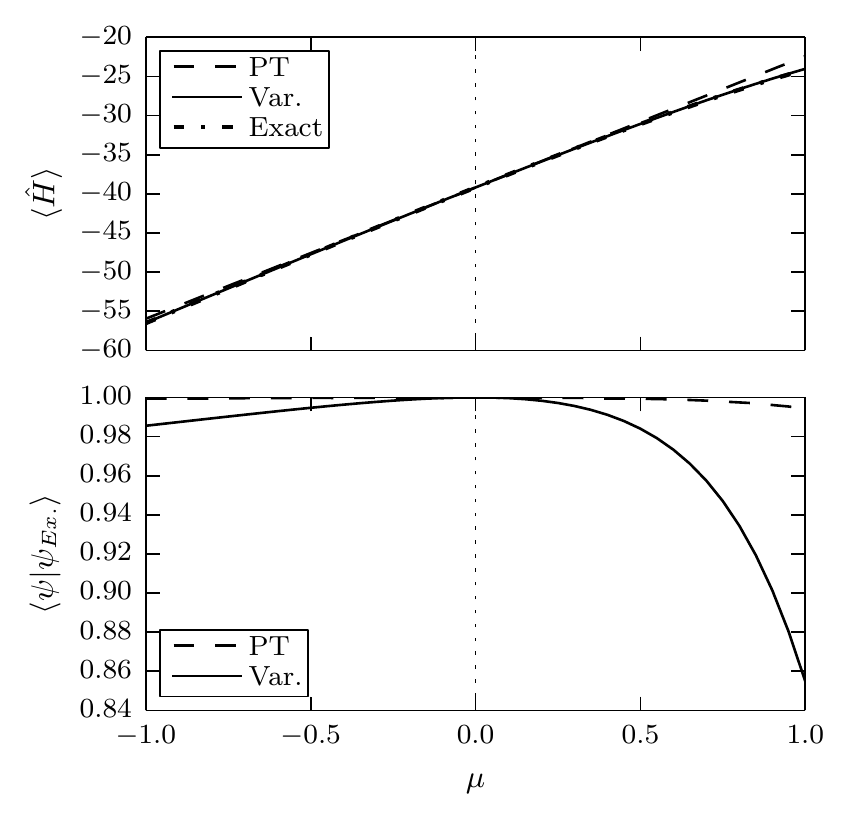}
\caption{Results for the inhomogeneous BCS model. \textbf{Top}: Variational energy, exact ground state energy, and first-order perturbation theory for different values of the perturbation strength. \textbf{Bottom}: Overlap of the exact ground state with the variational ground state and the ground state of the unperturbed model.  \label{fig:bcs:ener_ov}}
\end{figure}
\end{center}

The other emblematic example of Richardson-Gaudin models is the Richardson (or reduced BCS) Hamiltonian as given by 
\begin{equation}
\hat{H}_{BCS} = \sum_{i=1}^L \epsilon_i S_i^0 + g \sum_{i,j=1}^LS_i^{\dagger}S_{j}.
\end{equation}
This Hamiltonian can be used to describe fermion pairing in e.g. nuclear pairing and superconductivity \cite{talmi_simple_1993}, and is exactly solvable under the key assumption that the pairing interactions are uniform and fully determined by a single pairing constant $g$ \cite{richardson_restricted_1963,richardson_exact_1964}. Because of this exact solvability, this model has recently become a testing ground for novel many-body methods focusing on pairing interactions \cite{sambataro_treatment_2013,degroote_polynomial_2016,ripoche_combining_2017,gomez_attenuated_2017}. It's worthwhile to stress that the proposed integrability-based method will return the exact ground state energy of this model by construction. Moving away from integrability, the restriction of uniform interactions can be relaxed by introducing non-uniformities in a perturbative way, resulting in a more physical model. The Hamiltonians under consideration are of the form
\begin{equation}
\hat{H}_{BCS} = \sum_{i=1}^L \epsilon_i S_i^0 +  \sum_{i,j=1}^L G_{ij}S_i^{\dagger}S_{j}.
\end{equation}
While such models are solvable by $U(1)$-breaking BCS mean-field theory in the thermodynamic limit, it is important to obtain an accurate description for medium-size systems as well \cite{degroote_polynomial_2016,ripoche_combining_2017,gomez_attenuated_2017}. In fact, it has been shown that the Richardson-Gaudin equations are equivalent to the BCS mean-field equations for thermodynamically large systems, and as such the BCS wave function and the Bethe ansatz wave function coincide in this limit \cite{roman_large-n_2002}. The results are presented in Figure \ref{fig:bcs:ener_ov} for an interaction matrix $G_{ij} = g + \mu g_{ij}$, with $g=-1$ and $g_{ij}$ random numbers uniformly distributed over the interval $[0,1]$. We again take system size $L=12$, parameters according to the picket-fence model, and take $L=2N$ corresponding to half-filling.
\begin{figure*}[tb!]                 
 \begin{center}
 \includegraphics{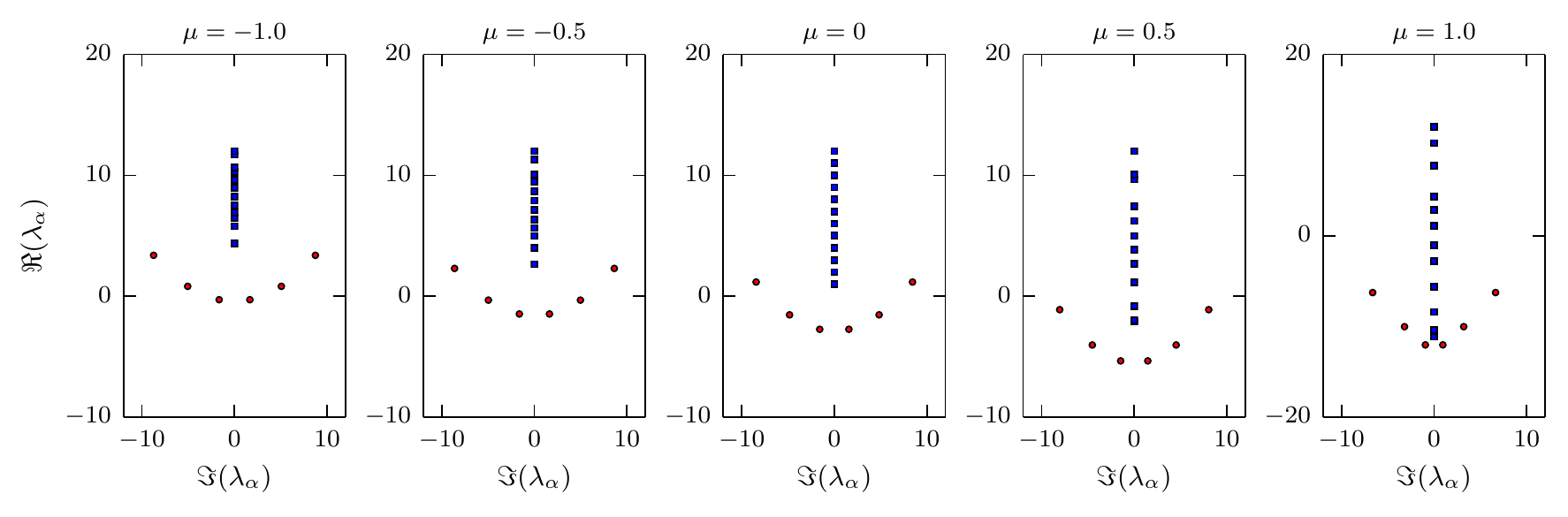} 
 \caption{Results for the inhomogeneous BCS model. Position of $\vec{\epsilon}$ (squares) and $\vec{\lambda}$ (dots) for the variationally optimized wave function in the complex plane at different values of the perturbation strength.\label{fig:rap_BCS}}
 \end{center}
\end{figure*}
The same behaviour as for the central spin model can be observed, where it should be noted that the error on the energy and overlap is much smaller compared to the results for the central spin model. This implies that a general pairing Hamiltonian can already be efficiently approximated by taking the average pairing interaction as single parameter, consistent with the success of BCS mean-field theory in the description of such Hamiltonians.  From the structure of the optimized wave function  (Figure \ref{fig:rap_BCS}), it can be seen that only minor modifications are necessary in order for the wave function to provide an accurate description.

\section{Conclusion and discussion}
\label{sec:concl}
In this work, we showed how the ground states of non-integrable Hamiltonians consisting of an integrable (Richardson-Gaudin) Hamiltonian and an integrability-breaking Hamiltonian can be approximated by modified eigenstates of related integrable Hamiltonians. Due to the inherent structure of these Bethe ansatz eigenstates, it is possible to efficiently calculate and minimize the expectation value of given Hamiltonians with respect to these states, and we showed how such a variational approach can be implemented. This was then shown to provide accurate results for select perturbed non-integrable Hamiltonians, where the accuracy of the variational approach is only limited by the appearance of avoided level crossings in the spectrum of non-integrable Hamiltonians. When the exact ground state can be considered a perturbation of the non-perturbed integrable Hamiltonian (i.e. there are no avoided crossings), the variational optimization starting from the non-perturbed ground state will provide accurate results. The effects of such crossings can then be taken into account by variationally optimizing excited states of the integrable Hamiltonian, instead of restricting the optimization to the ground state of the integrable model.

At present the selection of the proper excited state on which to perform the variational optimization is the main bottleneck in the procedure. One can envision several methods to cope with this problem. The method used in this paper is to capitalize on the physical insight in the perturbation. Often, the integrability-breaking term in the Hamiltonian itself has a clear physical interpretation, and it is only the competition between the integrable and non-integrable part of the Hamiltonian which is the main cause for complications. Consequently, the correct choice of variational manifold among the excited states can be deduced from the ground state structure of the integrability-breaking term. This is the approach used in the present paper, however other methods will be explored in the future, making use of ideas of stochastic sampling, the correspondence with Coupled-Cluster approaches \cite{limacher_new_2013,boguslawski_efficient_2014,boguslawski_efficient_2014,boguslawski_nonvariational_2014,limacher_simple_2014,henderson_seniority-based_2014, stein_seniority_2014,degroote_polynomial_2016} or the $pp$-TDA adiabatic connection \cite{de_baerdemacker_richardson-gaudin_2012,ring_nuclear_2004}.

\section*{Acknowledgements}
P.W.C. acknowledges support from a Ph.D. fellowship and a travel grant for a long stay abroad at the University of Amsterdam from the Research Foundation Flanders (FWO Vlaanderen). J.-S. C. acknowledges support from the Foundation for Fundamental Research on Matter (FOM) and from the Netherlands Organization for Scientific Research (NWO). This work forms part of the activities of the Delta-Institute for Theoretical Physics (D-ITP).

\appendix

\section{Expectation values from inner products}
\label{app:corr}
In this Appendix, we show how to obtain determinant expressions for expectation values starting from the inner product between two Bethe states. This construction is based on the commutation properties of the Gaudin algebra, and we will illustrate this for the expectation value of $S_i^z$. Then we have
\begin{align}
[S_i^0, S^{\dagger}(\lambda_{\alpha})] = \frac{S_i^{\dagger}}{\epsilon_i-\lambda_{\alpha}}, \qquad [[S_i^0, S^{\dagger}(\lambda_{\alpha})],S^{\dagger}(\lambda_{\beta})] = 0,
\end{align}
where we can rewrite the first commutator as 
\begin{equation}
[S_i^0, S^{\dagger}(\lambda_{\alpha})]  = \lim_{\lambda \to \epsilon_i} \frac{\epsilon_i-\lambda}{\epsilon_i-\lambda_{\alpha}} S^{\dagger}(\lambda).
\end{equation}
The action of $S_i^0$ on a Bethe state is then given by
\begin{align}
S_i^0 \ket{\lambda_1 \dots \lambda_N} &=  S_i^0 \prod_{\alpha=1}^N S^{\dagger}(\lambda_{\alpha})\ket{\downarrow \dots \downarrow} \nonumber\\
& = \sum_{\alpha=1}^N \prod_{\beta \neq \alpha}^N S^{\dagger}(\lambda_{\beta})[S_i^0, S^{\dagger}(\lambda_{\alpha})] \ket{\downarrow \dots \downarrow} \nonumber \\
& \qquad \qquad +\prod_{\alpha=1}^N S^{\dagger}(\lambda_{\alpha})S_i^0\ket{\downarrow \dots \downarrow} \nonumber \\
& =\lim_{\lambda \to \epsilon_i} \sum_{\alpha=1}^N \frac{\epsilon_i-\lambda}{\epsilon_i-\lambda_{\alpha}} S^{\dagger}(\lambda)  \prod_{\beta \neq \alpha}^N S^{\dagger}(\lambda_{\beta})\ket{\downarrow \dots \downarrow} \nonumber \\
& \qquad \qquad  -\frac{1}{2} \prod_{\alpha=1}^N S^{\dagger}(\lambda_{\alpha})\ket{\downarrow \dots \downarrow},
\end{align}
where the structure of Bethe states can again be recognized, with the variable $\lambda_{\alpha}$ replaced by $\lambda$, making them off-shell and allowing this to be rewritten as
\begin{align}
S_i^0 \ket{\lambda_1 \dots \lambda_N}& =\lim_{\lambda \to \epsilon_i} \sum_{\alpha=1}^N \frac{\epsilon_i-\lambda}{\epsilon_i-\lambda_{\alpha}} \ket{{\lambda_1} \dots \underset{\alpha}{\lambda} \dots {\lambda_N}} \nonumber \\
& \qquad \qquad  \qquad   -\frac{1}{2} \ket{\lambda_1 \dots \lambda_N}.
\end{align}
Expectation values now follow by taking the inner product of this state and $\ket{\lambda_1 \dots \lambda_N}$, and this can be expressed as a sum of determinants once we have an expression for 
\begin{equation}
\lim_{\lambda \to \epsilon_i}\frac{\epsilon_i-\lambda}{\epsilon_i-\lambda_{\alpha}}  \braket{\lambda_1 \dots \lambda_N | {\lambda_1} \dots \underset{\alpha}{\lambda} \dots {\lambda_N}},
\end{equation}
which follows from the known inner product of an on-shell state ($\ket{\lambda_1 \dots \lambda_N}$) with an off-shell state ($\ket{{\lambda_1} \dots \underset{\alpha}{\lambda} \dots {\lambda_N}}$). In many cases, this sum over determinants can even be further simplied, as shown in Refs. \onlinecite{zhou_superconducting_2002,links_algebraic_2003,faribault_exact_2008}. More specifically, for a single-spin operator, the number of determinants that need to be calculated equals $N$. For a two-spin operator $S_i^z S_j^z$ or $S_i^{\dagger}S_j$, the number of determinants will be given by $N^2$ when applying a similar commutator scheme\cite{zhou_superconducting_2002,links_algebraic_2003}. However, starting from the Slavnov determinant this summation can again be reduced to the evaluation of $2N$ (for $S_i^z S_j^z$) or $N$ (for $S_i^{\dagger}S_j$) determinants through some algebraic manipulations \cite{faribault_exact_2008}. For a given Hamiltonian containing $L_s$ ($L_d$) single-spin (double-spin) operators, the expected scaling of $\mathcal{O}(L_s N+L_d N^2)$ can hence be reduced to $\mathcal{O}(L_s N +L_d N)$. For the central spin model, this final expression results in a total number of determinants scaling as $\mathcal{O}(LN)$, whereas this number scales as $\mathcal{O}(L^2N)$ for the reduced BCS Hamiltonian.

\bibliography{MyLibrary.bib}
\end{document}